\documentclass[12pt, a4paper]{article}
\usepackage{geometry}
\usepackage[utf8]{inputenc}

\usepackage[T1]{fontenc} 

\usepackage{bm}
\usepackage{secdot}
\usepackage{mathptmx}
\usepackage{float}
\usepackage{soul}
\usepackage{authblk}

\RequirePackage{graphicx}
\RequirePackage{amsmath}

\title{Dissipative Unified Dark Fluid: Observational Constraints} 

\author[a]{Esraa Ali Elkhateeb\footnote{dr.esraali@sci.asu.edu.eg (corresponding author)}}
\author[b]{Mahmoud Hashim\footnote{mahmoud.hashim@bue.edu.eg}}

\affil[a]{Ain Shams University, Faculty of Science, Physics Department, Abbaseia, 11566, Cairo, Egypt.}
\affil[b]{Centre for Theoretical Physics, The British University in Egypt, P.O. Box 43, El Sherouk City, Cairo 11837, Egypt.}

\begin{document}

\maketitle

\abstract{We adopt a standard FRW cosmology with a unified scenario, where the usual dark matter and dark energy sectors are replaced by a single dissipative unified dark fluid (DUDF). The equation of state of such fluid can asymptote between two power laws. As a result, it enables fluid to have a smooth transition from dust at early times to dark energy at late times. The dissipation is represented by a bulk viscosity with a constant coefficient, whereas shear viscosity is excluded due to the isotropy of the universe. We performed a likelihood analysis using recent observational datasets from local $H_0$ measurements, Type Ia supernovae, observational Hubble data, baryon acoustic oscillations, and cosmic microwave background to put cosmological constraints on the model. The special case of the non-dissipative unified dark fluid (UDF) is also studied, while a similar analysis is performed on the $\Lambda$CDM model for comparison. We got an $H_0$ value of $70.02$ $km$ $s^{-1}  Mpc^{-1}$ for DUDF and $70.25$ $km$ $s^{-1}  Mpc^{-1}$ for the UDF model. our analyses revealed that between the three analyzed models, the DUDF has the lower $\chi_{\rm min}^2$-value. Based on model selection statistics in the form of the Akaike Information Criterion (AIC), we compare different models to select the favored one due to the observational data used. Our results revealed that the UDF model has the minimum AIC, with the conclusion that it is the most favorable model for the data. $\Delta AIC$ value of other models are then measured to this model. This difference indicated that the DUDF model is a substantial model on the level of empirical support. Additionally, we studied the evolution of the deceleration parameter, the effective equation of state parameter, and the density parameter. We also estimated a value for the viscosity of the cosmic fluid. We found that our unified fluid doesn't deviate from the standard $\Lambda$CDM model at early times, with the ability to play the role of the cosmological constant by accelerating the universe at the late times.}
 
\flushbottom

\section{Introduction}
\label{sec:intro}
The accelerating expansion of the universe was discovered in 1998 from supernovae data and confirmed by baryon acoustic oscillation (BAO) data and the analysis of galaxies clustering \cite{riss}-\cite{dos1}. The phenomenon was unexpected due to the gravitational attraction of normal matter in the universe and is attributed to some peculiar energy, termed dark energy. One explanation is through the cosmological constant, which was previously suggested by Einstein \cite{wien}. This idea was the base for the standard model of cosmology, the $\Lambda$CDM model, in which $\Lambda$ represents the cosmological constant and plays the role of dark energy, while CDM is the Cold Dark Matter, another sort of peculiar matter which is gravitating but non-baryonic.  According to recent cosmological observations, $95 \%$ of the total energy budget of the universe is composed of these dark sectors \cite{plnk18}.  However, the cosmological constant approach, although promising, suffers a conceptual problem \cite{book}, \cite{ombr}, in addition to the famous coincidence problem \cite{coprob}. Such problems have motivated the construction of alternative scenarios to the cosmological constant dark energy. In some of these scenarios, the dark energy is dynamic. In other scenarios the dark matter and dark energy are interacting \cite{khur}, \cite{pali}, while in some others they are non-interacting \cite{gzaho}, \cite{revera}.  There is also a class of models that considers unification between the two sectors. In these models, the two dark sectors are represented by a single dark fluid where the late transition from DM to DE can smoothly occur \cite{bosh}, \cite{wzim}. Chaplygin gas model and its generalizations are examples of these models \cite{chap1}, \cite{chap2}. There is also the new barotropic dark fluid model proposed by Elkhateeb \cite{viab}.

As viscosity is a physical property of fluids, cosmological viscosity attracted attention. Dissipative processes in cosmology have been discussed since the work of Zel'dovich \cite{zel} and Misner \cite{mis} in 1967. In his work in 1971, Weinberg \cite{wein} claims that vanishing bulk viscosity for a general dissipative fluid is just an exception rather than a rule. Many authors also discussed the viscous bulk pressure effects relating the cosmological particle production, e.g., \cite{bel75}-\cite{fran}. Viscous pressure inducing cosmic expansion also brings attention. A study made by Giovannini \cite{giov} declines the possibility of the existence of an accelerated phase driven solely by viscosity, where the tensor to scalar ratio exceeds the observational limit. He concluded that the viscous modes are only tolerable as a subleading component of a dominant adiabatic mode from a different physical origin. Some authors also discussed removing instabilities in the fluid power spectrum by treating viscosity as the physical origin for entropy perturbation, which in turn smoothes out anisotropies in the universe \cite{ent1}-\cite{ani}. Cosmological viscosity also included studying the early universe and entropy generation via cosmic inflation \cite{inf1}-\cite{inf2}, and bulk viscosity driven late universe acceleration \cite{late1}-\cite{late2}. Recently, Brevik and Normann \cite{sym} inferred symmetry between the early-time and present-day viscous cosmology by using a bulk viscosity function proportional to the Hubble parameter $H$. Victor et al. \cite{vic} discussed the bulk viscosity effects and matter production as an alternative for the description of the late times behavior of the observable universe. Alleviation of the $H_0$-tension between direct and indirect measurements is also a motivation for studying viscous models \cite{ten1}-\cite{ten2}. Many authors also studied The viscosity of the cosmic unified dark fluids \cite{uni1}-\cite{uni2}. Other aspects concerning dissipative cosmology can be found in \cite{diss1}-\cite{diss3}.

In this work, we aim to study the evolution of the universe under the effect of bulk viscosity using the latest cosmological observations. We adopt the new unified viscous dark fluid model introduced by Elkhateeb in a previous work \cite{diss4}. The dark fluid introduced has an equation of state with two power-law asymptotes, consequently can interpolate between dust and dark energy at early and late times. The added dissipative effects are in the form of bulk viscosity with a constant viscosity coefficient. Shear was considered negligible due to the isotropy of the universe. We implement recent observational datasets including local $H_0$ measurements, Type Ia Supernovae (SNe Ia), observational Hubble data (OHD), Baryon Acoustic Oscillations (BAO), and Cosmic Microwave Background (CMB) to perform robust analysis. The paper is organized as follows: Sec. 2 represents the theory of viscous fluid-based cosmology. Sec. 3 constrains some of the model parameters based on the asymptotic behavior of the model. In sec. 4, recent datasets used to validate the analysis procedure are presented. In sec. 5, we compare models based on model selection statistics in the form of the Akaike Information Criterion (AIC) to select the favorable model according to the observational data used. In sec. 6, we study the cosmological evolution due to our fluid model.
Sec. 7 for estimation of the present-day viscosity coefficient of the cosmic fluid.  We finally concluded in sec. 8.

\section{Fluid Based Cosmological Model} \label{sec1}
Einstein's equation for GR is given by
\begin{equation}
R_{\mu\nu}-\frac{1}{2}g_{\mu\nu}R=\frac{8 \pi G}{c^4} T_{\mu\nu} ,     
\label{field}
\end{equation}
where $R_{\mu\nu}$ is the Ricci curvature tensor, $R$ is the scalar curvature, $g_{\mu\nu}$ is the space metric and $T_{\mu\nu}$ is the 
energy-momentum tensor given by
\begin{equation}
T_{\mu\nu}=\rho U_\mu U_\nu + p h_{\mu\nu} ,
\label{Ttensor}
\end{equation}
where $\rho$ is the energy density of the cosmic fluid, $p$ is its pressure, and $h_{\mu\nu}=U_\mu U_\nu-g_{\mu\nu}$ is the projection tensor to the $3-$space orthogonal to the fluid element, where in comoving coordinates the four-velocity $U_\mu={\delta^0}_\mu$.

Considering units with $c=1$, the metric in the standard FRW cosmology is given by
\begin{equation}
ds^2=dt^2-a^2(t)(\frac{dr^2}{1-k r^2}+r^2 d\Omega^2) ,
\label{metric}
\end{equation}
where $a(t)$ represents the scale factor, $k$ is the curvature constant for the space with the three distinct values $-1$, $0$, and $1$, represent respectively a spatially open, flat, and a closed universe. $d \Omega^2 = d \theta^2 + sin^2 \theta d \phi^2$, with $\theta$ and $\phi$ are the usual polar and azimuthal angles of spherical coordinates. The coordinates $(t, r, \theta, \phi)$ are the comoving coordinates.
Considering our universe is filled with several components: radiation, baryons, and dark fluid, each with pressure $p_i$ and energy density $\rho_i$, Einstein's equation (\ref{field}) together with the metric (\ref{metric}) will lead to the Friedmann equations
  
\begin{align}
\frac{\dot{a}^2}{a^2}=\frac{1}{3} \sum_i \rho_i	\label{hsq} ,\\
\frac{\ddot{a}}{a}=-\frac{1}{6} \sum_i (\rho_i + 3 p_i) , \label{hddot}  
\end{align}
where we consider flat geometry and units with $8 \pi G=1$. The conservation equation $T^{\mu}_{\nu(i);\mu}=0$ gives
\begin{equation}
\dot{\rho_i}+3 \frac{\dot{a}}{a} (\rho_i + p_i)=0 ,
\label{cons}
\end{equation}
where $T^{\mu}_{\nu(i)}$ is the energy momentum tensors for the $i-th$ component.

The effective pressure of dark fluid component $p_{\rm fld}$ is considered to consist of an adiabatic part $p_a$ plus a viscous term 
\begin{equation}
  p_{\rm fld}=p_a-\vartheta \zeta(\rho) ,
\label{pres}
\end{equation}
where $\vartheta= {U^\mu}_{;\mu}=3 H$ is the expansion rate of the fluid with $H=\frac{\dot{a}}{a}$ is the Hubble parameter, $\zeta=\zeta(\rho)$ is the coefficient of bulk viscosity that arises in the fluid and is restricted to be positive. For the adiabatic part, $p_a$, we adopt a barotropic pressure that can asymptote between dust at the early time and DE at the late time \cite{viab}. It has the form
\begin{equation}
  p_a=-\rho_{\rm fld} + \frac{\gamma \rho^n_{\rm fld}}{1+\delta \rho^{m}_{\rm fld}} ,
\label{van1}
\end{equation}
where $\gamma$, $\delta$, $n$, and $m$ are free parameters. As mentioned in \cite{viab}, this form of adiabatic pressure has the advantage that it enables interpolation between different powers for the density, which allows for smooth phase transitions during the universe evolution. It also has the advantage of having a general equation of state (EoS) for DE that enables the cosmological constant as special case. The bulk viscosity coefficient is considered to take the form
\begin{equation}
\zeta(\rho_{\rm fld}) = \zeta_0 \rho^{\nu}_{\rm fld} ,
\label{vis}
\end{equation}
where $\zeta_0$ and $\nu$ are constants. In this work we adopt a constant bulk viscosity coefficient $\zeta(\rho) = \zeta_0$, so that $\nu=0$. This assumption is attractive due to its mathematical simplicity while it is physically acceptable. Accordingly, our viscous dark fluid has the effective equation of state

\begin{equation}
p_{\rm fld} = - \rho_{\rm fld} + \frac{\gamma \rho_{\rm fld}^n}{1+\delta \rho_{\rm fld}^m} - 3 \zeta_0 H .
\label{eqos}
\end{equation}

\section{The Model Parameters}                             
The dark sector of the universe is supposed to consist of viscous dark fluid with an EoS asymptotes between two power laws characterizing dust at early time and dark energy at late time. The asymptotic behavior of the model can shed light on the relationship between its parameters, which reduces the number of degrees of freedom of the model. In doing so, we define the following dimensionless parameters,
\begin{equation}
\tilde{\rho}_{\rm fld}=\delta^{\frac{1}{m}} \rho_{\rm fld} ,
\label{dimlesrho}
\end{equation}
and
\begin{equation}
\tau=\frac{\gamma}{{\delta}^{\frac{(2n-1)}{{2m}}}} t ,
\label{dimlestau}
\end{equation}
so that $\left[\delta^{\frac{1}{m}}\right] = \left[\rho^{-1}\right] = \left[H^{-2}\right]$. Accordingly, the conservation equation for the fluid reduces to
\begin{equation}
\frac{d \tilde{\rho}_{\rm fld}}{d \tau}= -\sqrt{3}\left(\frac{{\tilde{\rho}_{\rm fld}}^{n+\frac{1}{2}}}{1+{\tilde{\rho}}^{m}_{\rm fld}} - \tilde{\zeta_0} \tilde{\rho}_{\rm fld} \right) ,
\label{consw}
\end{equation}
with 
\begin{equation}
\tilde{\zeta_0}= \frac{\sqrt{3}}{\gamma} \delta^{\frac{(2n-1)}{{2m}}} \zeta_0 ,
\label{dimleszet}
\end{equation}
is also dimensionless, while $\left[\zeta_0\right] = \left[H\right]$. 

DUDF has to mimic dust for large $\tilde{\rho}$ and DE for small $\tilde{\rho}$, which means that $d \tilde{\rho}/d \tau$ will be in proportion to $\tilde{\rho}$ for large $\tilde{\rho}$ and tends to zero for small $\tilde{\rho}$. Accordingly, the parameter $n$ has to be positive. Restricting $m$ also to be positive, relation (\ref{consw}) tends asymptotically to the following form for large $\tilde{\rho}$
\begin{equation}
\frac{d \tilde{\rho}}{d \tau} \sim -\sqrt{3}\left(\tilde{\rho}^{(n-m+\frac{1}{2})} - \tilde{\zeta_0} \tilde{\rho} \right) .
\label{conslarge1}
\end{equation}

For small $\tilde{\zeta_0}$, the first term in the above equation dominates for $(n-m) \geq 1$. Under this condition, Eq. (\ref{conslarge1}) reduces to
\begin{equation}
\frac{d \tilde{\rho}}{d \tau} \sim -\sqrt{3} \; \tilde{\rho}^{(n-m+\frac{1}{2})} .
\label{asylarge}
\end{equation}
Comparing this with Friedmann equation for dust, namely
\begin{equation}
\frac{d \tilde{\rho}}{d \tau}= -\sqrt{3} \; \frac{\delta^{\frac{(n-1)}{m}}}{\gamma} \; \tilde{\rho}^{\frac{3}{2}} ,
\label{conslarge}
\end{equation}
we directly get
\begin{equation}
n=m+1 \; , \; \; \; \; \; \; \; \gamma = \delta .
\label{pars}
\end{equation}
As a result
\begin{equation}
\tilde{\zeta_0}= {\sqrt{\frac{3}{\tilde{\delta}}}} \; \zeta_0 ,
\label{dimleszet1}
\end{equation}
with
\begin{equation}
\tilde{\delta} = \delta^{-\frac{1}{m}},
\label{delta}
\end{equation}
has the dimensions of $H^2$; $\left[\tilde{\delta}\right] = \left[\rho\right] = \left[H^2\right]$.

In a previous work \cite{diss4}, we introduced the fundamental properties of the DUDF model and estimated its parameters using observations of different cosmological phenomena such as today's value of deceleration parameter, the redshift value of deceleration-acceleration transition, and the age of the universe. In this work, we seek to use the available cosmological datasets to constrain the model parameters robustly.

\section{Data Analysis}
\subsection{Observational Data}                         
We now aim to constrain the DUDF model using observational data. We relied on the recent observational datasets from local $H_0$ measurements, Type Ia Supernovae (SNe Ia), observational Hubble data (OHD), Baryon Acoustic Oscillations (BAO), and Cosmic Microwave Background (CMB).

In our analysis, the cosmological model is specified by the total relative density,
\begin{equation}
\Omega_{\rm tot}(z)=\Omega_{\rm r}(z) + \Omega_{\rm b}(z) + \Omega_{\rm fld}(z) .
\label{tot}
\end{equation}
The radiation term $\Omega_{\rm r}$ considers the standard thermal history into account with today's value $\Omega_{\rm r0}$ given by \cite{bing}
\begin{equation}
\Omega_{\rm {r0}}(z)=\Omega_{\gamma}\left[1 + \frac{7}{8} \left(\frac{4}{11}\right)^{4/3} N_{\rm eff}\right] .
\label{r0}
\end{equation}
where $\Omega_{\gamma} = 2.47298 \times {10}^{-5}/h^{-2}$ is the current value of photon density, $h$ is the dimensionless Hubble constant defined through $H_0=100 \; h \; km \; s^{-1} Mpc^{-1}$, and $N_{\rm eff} = 3.046$ is the effective number of neutrino species \cite{neff}.
 
\subsubsection{SNe Ia}
Type Ia supernovae (SNe Ia) have a known brightness whence can serve as standard candles and can be used for measuring cosmological distances. Pantheon Sample \cite{pan} is the most complete updated compilation of SNe Ia. Confirmed 276 SNe Ia from Pan-STARRS1 Medium Deep Survey at $0.03 < z < 0.65$ are combined with previous available SNe Ia samples from other surveys to form a sample of $1048$ SNe Ia spanning the redshift range $0.01 < z < 2.3$. Data for the apparent magnitude $m(z)$ is given instead of the distance modulus $\mu$ as the absolute magnitude $M$ of SNe Ia degenerates with $H_0$. Observations of the apparent magnitude of SNe Ia, $m(z)$, are related to the distance modulus $\mu(z)$ through the relation
\begin{equation}
\mu(z)=m(z)-M=5 log_{10}(\frac{D_L(z)}{10 \; pc}) ,
\label{mu}
\end{equation}
where the absolute magnitude $M$ is considered constant for all supernovae of Type Ia due to the standard candle hypothesis, and $D_L(z)$ is the luminosity distance given by
\begin{equation}
D_L(z)= D_H\, (1+z) \int^{z}_{0} \frac{d{z'}}{E({z'})} ,
\label{dl}
\end{equation}
with
\begin{equation}
D_H= \frac{c}{H_0} ,
\label{DH}
\end{equation}
is the Hubble distance, and the function $E(z)$ is given by
\begin{equation}
E(z)= \frac{H(z)}{H_0} .
\label{Ez}
\end{equation}

\subsubsection{Observational Hubble Data}
In addition to the SNe Ia observations, we also considered observational Hubble data (OHD) from CC and BAO measurements. CC techniques \cite{cctec} allow direct information about Hubble function as they are probe for direct measure of the differential age, $\frac{\Delta z}{\Delta t}$, of passively evolving galaxies, whence $H(z)=-\frac{1}{1+z}\frac{\Delta z}{\Delta t}$. Accordingly, They allow for model-independent measurements of Hubble function. However, CC is a good tool for Hubble function measure at redshifts $\leq 2$ \cite{chron}.

In this work, we considered the updated list compiled by Maga$\widetilde{n}$a et al. \cite{mag} which contains 31 points from CC measurements plus 20 points from clustering. However, we excluded the points of Alam et al. \cite{alam} as they are used in the BAO dataset.

\subsubsection{Local $H_0$ Measurements}
Additionally, we included in our analysis the local value of Hubble parameter $H_0$ provided by Riess et al. 2020 (R20) results \cite{ries20}. 

\subsubsection{BAO}
Anisotropic BAO analysis can measure the Hubble function $H(z)$ and comoving angular diameter distance $D_M(z)$ due to the BAO shifts perpendicular and parallel to the line of sight. The comoving angular diameter distance $D_M(z)$ for a flat universe is defined to be
\begin{equation}
D_M(z)= D_H \int^{z}_{0} \frac{d{z'}}{E({z'})} .
\label{dm}
\end{equation}
As changes to the pre-recombination energy density alter the radius of the sound horizon, BAO measurements really constrain the combinations $D_M(z)/r_d$ and $H(z) r_d$, where $r_d$ is the radius of sound horizon at the drag epoch
\begin{equation}
    r_d = \frac{c}{H_0} \int_0^{1/\left(1+z_d\right)}{\frac{da}{a^2\; E(a)\; \sqrt{3 \left(1 + \frac{3 \omega_b}{4 \omega_\gamma}\right)\; a}}}
    \label{rd}
\end{equation}
with $z_d$ is the redshift of the drag epoch, $\omega_b = \Omega_b h^2$ is the current physical density of baryons, and $\omega_\gamma = \Omega_\gamma h^2$ is the current physical density of photons.  To calculate the redshift at the drag epoch we considered the recent formula fitted through the machine learning approach by Aizpuru et al \cite{aiz}
\begin{equation}
    z_d = \frac{1 + 428.169 \;  \omega_b^{.256459} \;  \omega_m^{.616388} + 925.56 \;  \omega_{m}^{.751615}}{\omega_{m}^{.714129}}
    \label{zd}
\end{equation}
which is accurate up to $ \sim 0.001 \%$, where $\omega_m = \Omega_m h^2$ is the current physical density of matter.

We used the results of BAO measurements from Sloan Digital Sky Survey (SDSS-III) provided by  Alam et al. \cite{alam} for $D_M(z)/r_d$ and $H(z)  r_d$ at redshifts $z= 0.38$ , $0.51$, and $0.61$. The $\chi^2_{BAO}$ for these data is constructed as
\begin{equation}
   \chi^2_{BAO} = \left( \Bar{D}_{obs} - \Bar{D}_{th}\right)^T \; C_{SDSS}^{-1} \; \left( \Bar{D}_{obs} - \Bar{D}_{th}\right)
    \label{chiSDSS}
\end{equation}
where $\Bar{D}_{obs}$ is an abbreviation for 
\begin{equation}
    D_M(z) \frac{r_d^{fid}}{r_d} \, \, \, \, \, \,  \& \,\, \, \, \, \,  H(z) \frac{r_d}{r_d^{fid}}
\end{equation}
with $r_d^{fid} = 147.78$ Mpc. $ C_{SDSS}^{-1}$ is the inverse covariant matrix of the data. Observational data and covariant matrix are available publicly at the SDSS-III website \cite{data}.

\subsubsection{CMB}
CMB data provide information about the entire history of the universe up to the last scattering surface. In our analysis we use results of the compressed likelihood of the CMB power spectrum given in Chen et al. \cite{chen} based on the Planck-2018 TT,TE,EE+lowE data release. We use the results of ${\mathcal{R}, l_A, \omega_b}$ to constrain our model parameters. Here
\begin{equation}
\mathcal{R} =\frac{1}{c} D_M(z_\ast) \sqrt{\Omega_{m} H^{2}_{0}}
\label{shift}
\end{equation}
is the CMB shift parameter, and
\begin{equation}
    l_A = (1+z_\ast) \frac{\pi \; D_A(z_\ast)}{r_s(z_\ast)}
    \label{lac}
\end{equation}
is the acoustic scale at last scattering, with $z_\ast$ is the redshift at the photon decoupling epoch calculated using the recent formula of Aizpuru et al \cite{aiz}
\begin{equation}
z_\ast = \frac{391.672 \;  \omega_m^{-.372296} + 937.422 \; \omega_b^{-.97966}}{\omega_m^{-.0192951} \; \omega_b^{-.93681}} + \omega_m^{-.731631}
\label{zstar}
\end{equation}
which is accurate up to $ \sim 0.0005 \%$, and $r_s(z_\ast) = r_\ast$ is the comoving size of the sound horizon at $z_\ast$ given by
\begin{equation}
    r_\ast = \frac{c}{H_0} \int_0^{1/\left(1+z_\ast\right)}{\frac{da}{a^2\; E(a)\; \sqrt{3 \left(1 + \frac{3 \omega_b}{4 \omega_\gamma}\right)\; a}}}
    \label{rs}
\end{equation}

The $\chi^2$ for the CMB data is constructed as
\begin{equation}
    \chi^2_{CMB} = X^T \; C_{CMB}^{-1} \; X
    \label{chicmb}
\end{equation}
where
\begin{equation}
    X = 
    \begin{pmatrix}
        \mathcal{R} - 1.7502  \\
        l_A - 301.471 \\
        \omega_b - 0.02236
    \end{pmatrix}
    \label{xmat}
\end{equation}
and $C_{CMB}^{-1}$ is the inverse covariant matrix of the CMB data, where
\begin{equation}
   \left( C_{CMB} \right)_{ij} = \sigma_i \sigma_j D_{ij}
   \label{ccmb}
\end{equation}
with $D$ is the correlation matrix given by \cite{chen}
\begin{equation}
    D = 
    \begin{pmatrix}
        1.0 & 0.46 & -0.66 \\
        0.46 & 1.0 & -0.33 \\
        -0.6 & -0.33 & 1.0
    \end{pmatrix}
\end{equation}


\subsection{Parameters Estimation}
The probability distribution function, posterior, $p\left(\bm { \Theta| D}\right)$ for the model parameters $\bm{\Theta}$ given different observational datasets $\bm{D}$ is given by Bayes' theorem:
\begin{equation}
p\left(\bm{\Theta| D}\right) = \frac{p\left(\bm{D|\Theta}\right) p\left(\bm{\Theta}\right)}{p(\bm{D})} ,
\label{prob}
\end{equation}
where $p(\bm{\Theta})$ is the parameters prior, and $p(\bm{D})$, known as the evidence, is the probability of observing the data. The evidence $p(\bm{D})$ is constant and can be set equal to $1$ since, after all, we got the observations. The probability $p\left(\bm{D | \Theta}\right)$ is the expected distribution of data given the parameters, likelihood, and is given by:
\begin{equation}
\mathcal{L}= \prod^{i}_{N}\frac{1}{\sqrt{2\pi\sigma^{2}_{i}}} \exp(-\frac{1}{2} \chi^{2}_{\rm tot}) ,
\label{lik}
\end{equation}
where
\begin{equation}
\chi^{2}_{\rm tot}= \sum_{\rm d}\chi^{2}_{\rm d} ,          
\label{chitot}
\end{equation}
with d represents each data group.

To constrain DUDF model we used  the public Markov Chain Monte Carlo (MCMC) emcee code \cite{mcmc}. The parameter space were chosen to be seven-dimensional; the three basic cosmological parameters: Hubble constant $H_0$, today's baryon density $\Omega_{b}h^2$, and today's dark matter density $\Omega_{c}h^2$, plus the three model parameters: $m$, $\widetilde{\delta}$, and $\zeta_0$, in addition to the nuisance parameter represents the absolute magnitude $M$, i.e, $\bm{\Theta} = \{H_0, \Omega_{b}h^2, \Omega_{c}h^2, m, \widetilde{\delta}, \zeta_0, M\}$. The posterior is constructed from the likelihood and a flat prior on all parameters. In table \ref{prior} we summarized the priors of our parameters space.

\begin{table} [h!]
	\begin{center}
	\caption{Priors of the model parameters.}
		\begin{tabular} {| c | c | c | c | c | c | c |}
		\hline
		 $H_0$ & $\Omega_{b}h^2$  & $\Omega_{c}h^2$ & m & $10^{-4} \widetilde{\delta}$ & $\zeta_0$ & M \\  \hline
		  [65, 80] & [0.015, 0.028] & [0.1, 0.13] & [0.5, 2.5] & [1.3, 2.7] & [0, 6] & [-20, -18]  \\ \hline
		\end{tabular}
	\label{prior}
	\end{center}
\end{table}

\begin{figure} [h!]
	\centering
		\includegraphics[width=0.9 \textwidth] {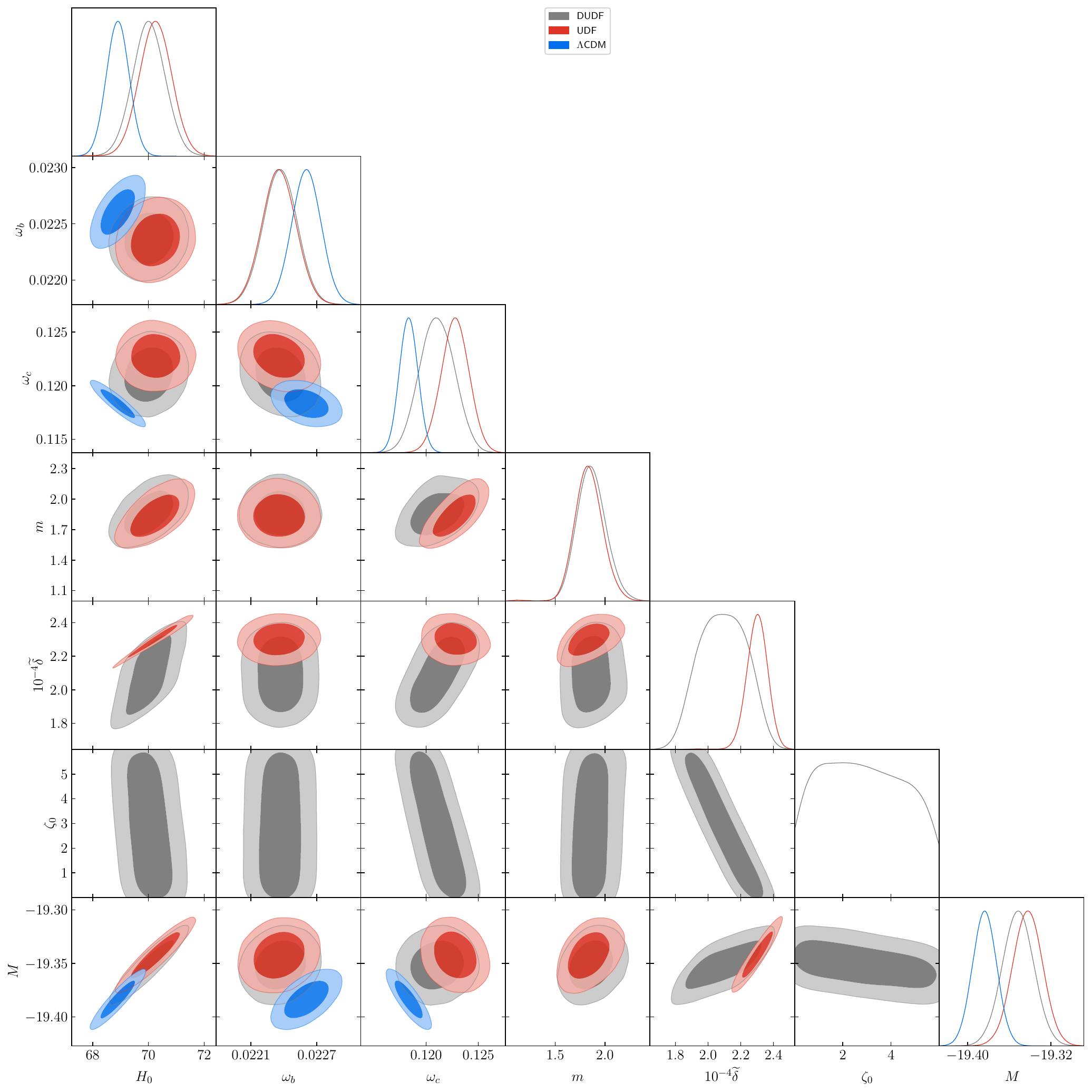}
	\caption{The marginalized $68\%$ and $95\%$ C.L. contours of the parameters of the three models. Shown also the one dimensional marginalized posterior distributions for the parameters.}
	\label {contour1}
\end{figure}

The previously mentioned five different datasets are used to constrain the parameters of our dissipative unified dark fluid, DUDF, in addition to the particular case of non-dissipative unified dark fluid, UDF. Similar analysis is also performed on the $\Lambda$CDM model for comparison. In Fig. \ref{contour1} we plot the $68\%$ and $95\%$ two-dimensional contours for the three models. In the Fig. we also show the one-dimensional marginalized posterior distribution for the parameters of the models.

From the Fig. we can notice the correlation between the different parameters. In particular, there is a correlation between $H_0$ and the dark fluid parameters $m$ and $\tilde{\delta}$. There is an anticorrelation between the parameters $\zeta_0$ and  $\tilde{\delta}$ too. One can also notice that the viscosity parameter $\zeta_0$ is poorly constrained by the data. The parameter $\tilde{\delta}$  in the DUDF model is poorly constrained, as well, due to its degeneracy with $\zeta_0$. In table \ref{tabmeans} we list the $68\%$ C.L. of the parameters of the three models under consideration.

\begin{table} [h!]
  \begin{center}
	\caption{$68\%$ C.L. parameters of cosmological models.}
	  \scalebox{0.7} {
		\begin{tabular} {|c| c | c | c |} 
		\hline 
		
		 Parameter & DUDF & UDF & $\Lambda$CDM \\ \hline
	
		 $H_0$ & $ 70.02\pm 0.56 $ & $ 70.25\pm 0.57 $ & $ 68.90\pm 0.39 $ \\ 
		      
		$\Omega_{b}h^2$ &  $ 0.022368\pm 0.00015 $ & $ 0.022356\pm 0.00015 $ & $ 0.02261\pm 0.00013 $ \\ 
		      
		$\Omega_{c}h^2$ &  $ 0.1211\pm 0.0016 $ & $ 0.1227\pm 0.0013 $ & $ 0.11832\pm 0.00086 $  \\ 
		      
		$m$ &  $ 1.87^{+0.13}_{-0.15} $ & $ 1.84\pm 0.14 $ & $--$ \\ 

		$10^{-4}$ $\widetilde{\delta}$ & $ 2.091\pm 0.13 $ & $ 2.299\pm 0.063 $ & $--$  \\
		
		$\zeta_0$ & $ 2.9^{+1.6}_{-2.3} $ & $ 0.0 $ & $--$  \\
		      
		$M$ &  $ -19.352\pm 0.015 $ & $ -19.343\pm 0.014 $ & $ -19.384\pm 0.011 $ \\ \hline
		\end{tabular} }
	\label{tabmeans}
	\end{center}
\end{table}

\section{Model Comparison}
In table \ref{chi} we display the $\chi^2_{\rm min}$-values for models under consideration. We can see that between the three models, the DUDF model has the lower value of the $\chi^2_{\rm min}$.

\begin{table} [h!]
	\begin{center}
	\caption{Summary of the information criteria results}
		\begin{tabular} {| c | c c c |}
		\hline
		  & DUDF  & UDF & $\Lambda$CDM  \\ \hline 
		 $\chi^2_{\rm min}$ & $ 1059.68 $ & $ 1059.78 $ & $ 1065.96 $ \\
		${AIC}$ &  $ 1073.68 $ & $ 1071.78 $ & $ 1073.96 $ \\
		$\Delta$ ${AIC}$ & $1.9$ &  $0$ & $2.2$ \\ \hline
		\end{tabular} 
	\label{chi}
	\end{center}
\end{table}

The $\chi^2$-statistics enables the determination of the best-fit parameters of a model. Nevertheless, they are not sufficient for deciding which model is the best, especially if the models have a different number of parameters. Many model selection statistical methods are proposed in the context of cosmology in the literature. The most commonly used is the Akaike information criteria (AIC) \cite{Akke}. It is more robust as it considers the number of parameters of each model.

The AIC is defined for a given model as
\begin{equation}
AIC = -2 ln \mathcal{L} + 2 p ,
\label{aic}
\end {equation}
where $p$ is the number of free parameters of the model. The model which minimizes the AIC is considered the better model. As a result, the second term in the criteria imposes a penalty against the extra number of parameters. In general, the AIC criterion is inclined to select the model that better fits the data \cite{park}. In the second row of table \ref{chi} we display the values of $AIC$ for the three models at hand. We can note that between the three models, the UDF model has the lower value of $AIC$. We can then conclude that the UDF model is the most favored by the data.

However, the difference between AICs of different models is that making sense. According to \cite{Burn}, if $\Delta_i = AIC_i - AIC_{\rm min}$, then model$_i$ which has $\Delta_i$ between 0 and 2 is considered as substantial on the level of empirical support, while models with $\Delta_i$ between 4 and 7 has considerably less support. For $\Delta_i > 10$, model$_i$ may be omitted in future consideration.

As the UDF model minimizes the $AIC$ value, the $\Delta_i$'s for other models are measured to this model. The results are included in the last row of table \ref{chi}. We can see from the table that $\Delta AIC$ for the DUDF model is $<2$. Following \cite{Burn}, we conclude that the DUDF model shows a substantial level of empirical support and is not worth more than a bare mention of evidence against the UDF model.

\section{Cosmological Evolution of the Model}
In this section, we study the cosmological evolution due to our fluid model and discuss the consistency of the model with observations.

\subsection{The Hubble Parameter}
\begin{figure} [h!]
	\centering
		\includegraphics[width=0.7\textwidth]{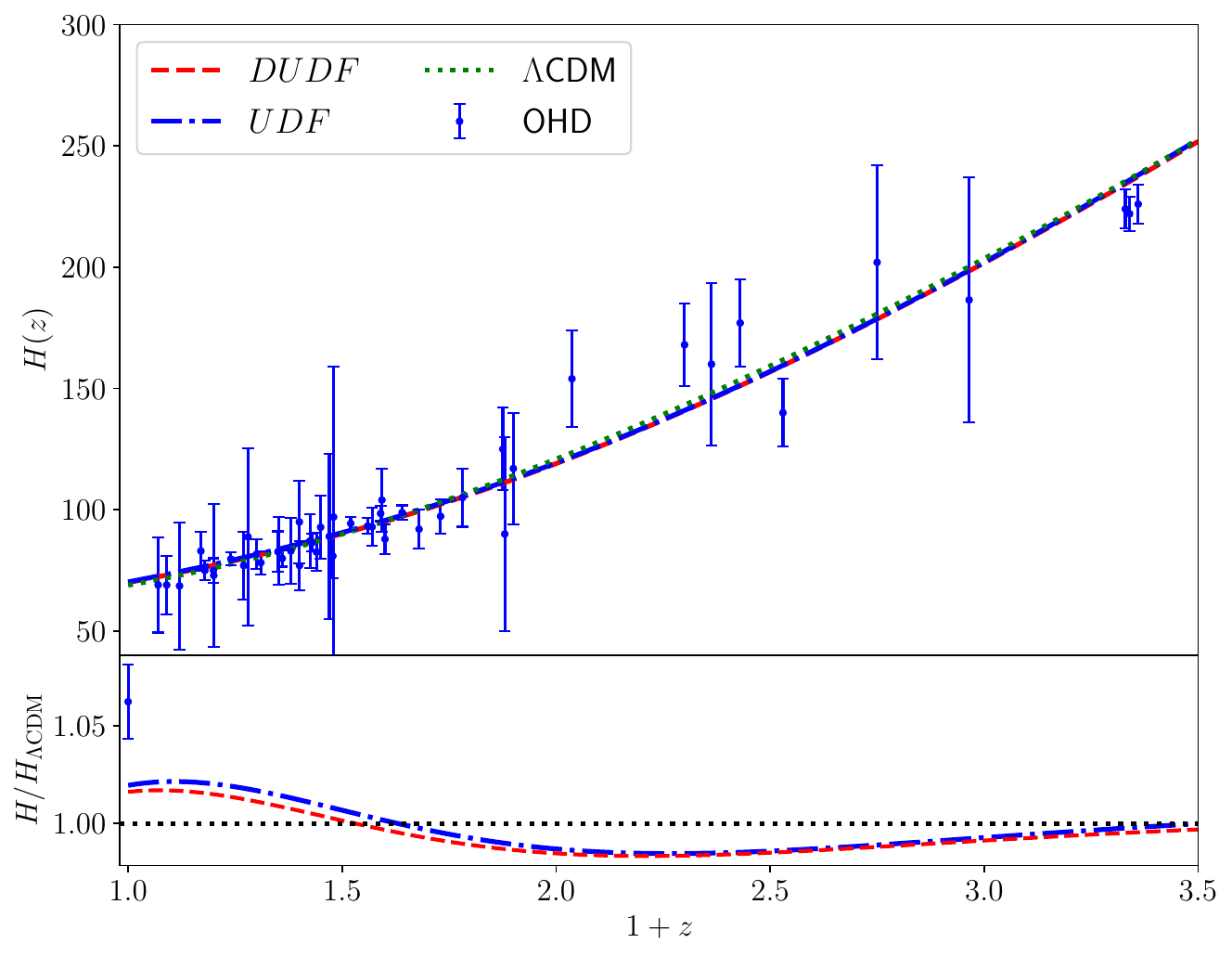}
	\caption{Upper panel: evolution of the Hubble parameter. Lower panel: Ratio between Hubble parameter of the fluid models to that of the $\Lambda$CDM model.}
	\label {hubble}
\end{figure}

One of the most important physical parameters in cosmology is the Hubble parameter. It describes the expansion history of the universe. In the upper panel of Fig. \ref{hubble} we plot the evolution of the Hubble parameter due to DUDF and UFD models against the OHD. We also displayed results from the $\Lambda$CDM model for comparison. In the lower panel of the figure, we present the relative ratio $H(z)/H_{\Lambda CDM}$.

The graph shows that the model represents the observational data very well. From the lower panel, we can see that our fluid follows the history of the standard cosmology at early times. As time passes, its deviation from the $\Lambda$CDM model is in the range of $\approx 2\%$.

\subsection{Supernovae SNe Ia Observations}
\begin{figure} [h!] 
	\centering
		\includegraphics[width=0.7\textwidth]{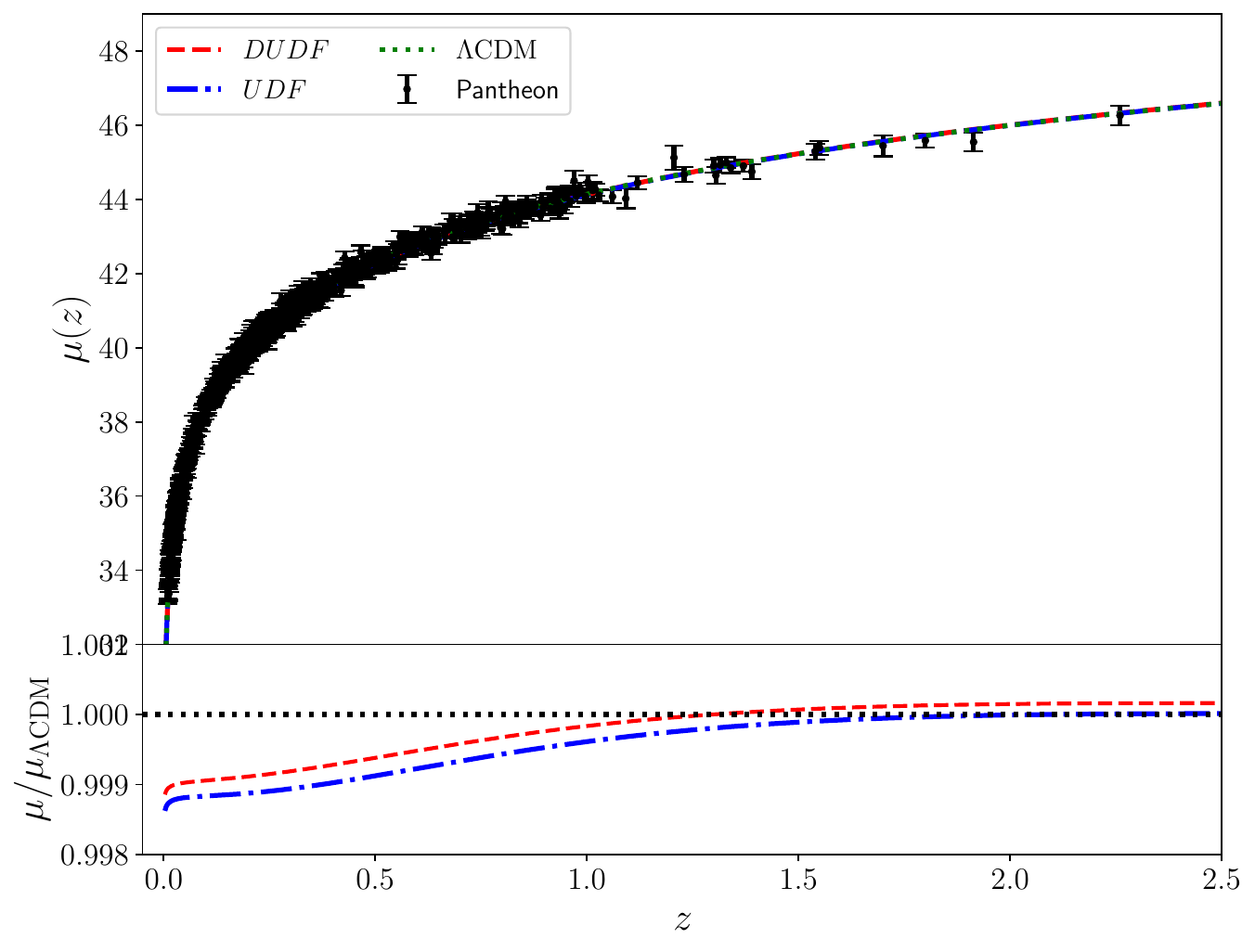}
	\caption{Upper panel: distance modulus from different models vs Pantheon data. Lower panel: ratio between the distance modulus from the fluid models to that of the $\Lambda$CDM model.}
	\label {SN}
\end{figure}

Distance measurements using SNe Ia lead to the discovery of the accelerating expansion of the universe. This was first discovered in $1998$. SNe Ia are referred to as standard candles, and their different characteristics extend to their stellar age distribution. The upper panel of Fig. \ref{SN} represents the distance modulus calculated from the fluid models, DUDF and UDF, against observations from Pantheon data. Results from the $\Lambda$CDM model are also displayed for comparison. The lower panel of the figure shows the ratio of the distance modulus from each of the fluid models to that of the $\Lambda$CDM model. We can see from the Fig. that the fluid models excellently fit the SNe Ia data and highly follow the $\Lambda$CDM results, where the deviations are of the order of $0.15 \%$.

\subsection{BAO Observations}
\begin{figure} [h!] 
	\centering
		\includegraphics[width=0.7\textwidth]{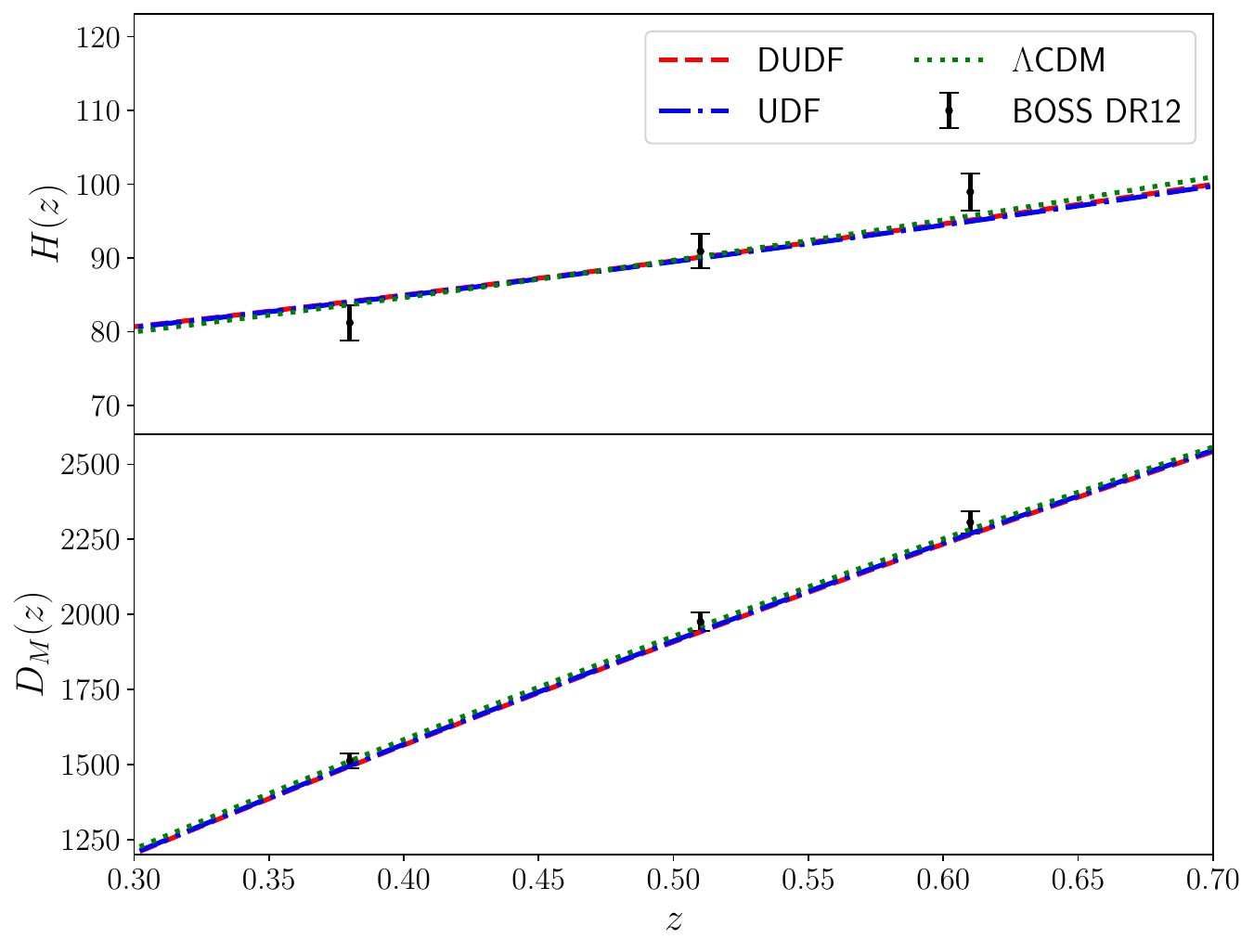}
	\caption{A comparison between the observational BAO data and theoretical calculations from the three models at hand.}
	\label {BAOfig}
\end{figure}

BAO matter clustering provides a standard ruler for length scale and can be used to measure the expansion history of the universe. In Fig. \ref{BAOfig} we compare theoretical calculations for Hubble function and comoving angular diameter distance from the three models at hand with the SDSS-III data. The Fig. shows that both DUDF and UDF models well represent the data.

\subsection{CMB Observations}
\begin{figure} [h!] 
	\centering
		\includegraphics[width=0.7\textwidth]{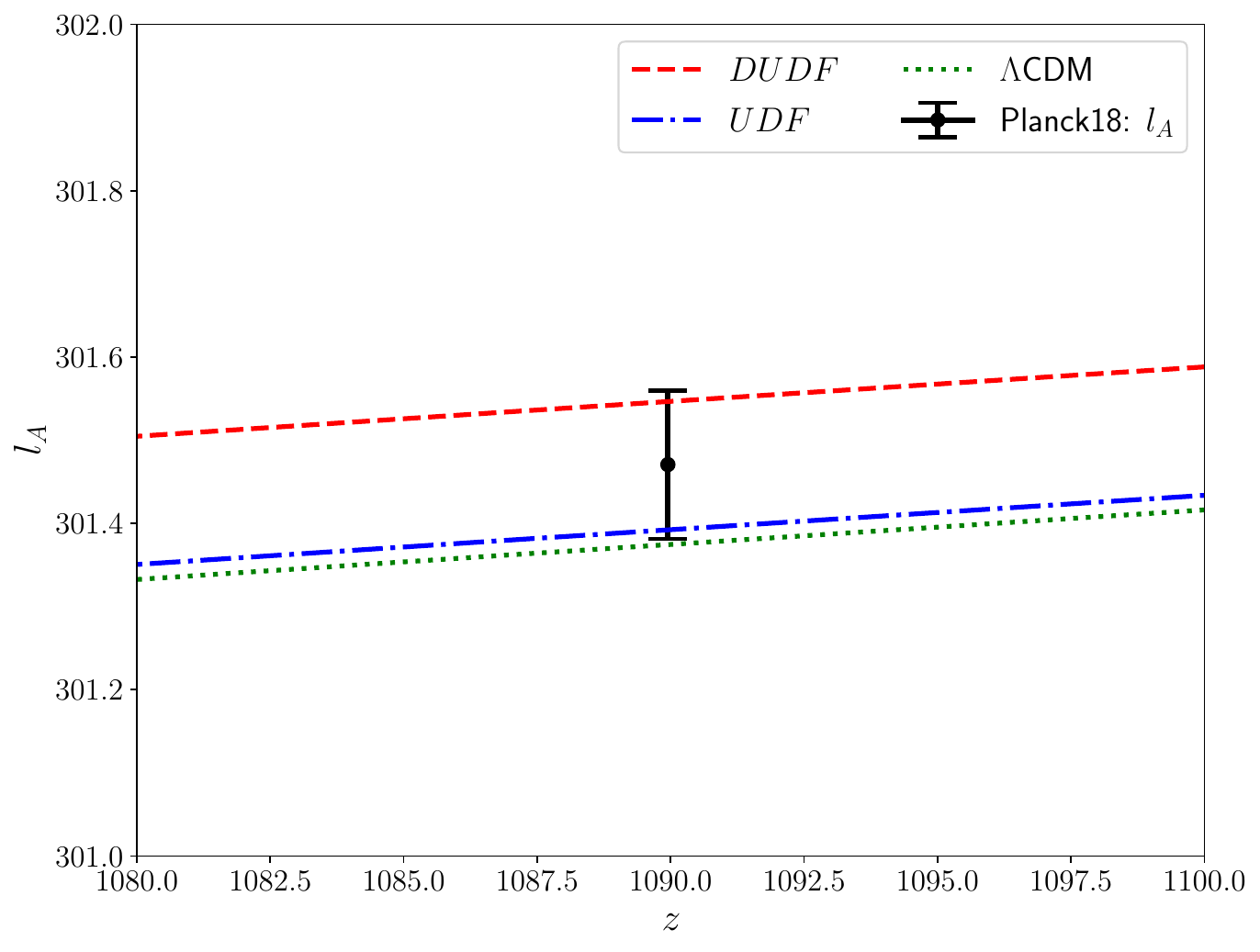}
	\caption{A comparison between the observational $l_A$ from Planck18 and the theoretical calculations from the three models at hand.}
	\label {cmbdta}
\end{figure}

The compressed likelihood of Planck data contains three quantities: the CMB shift parameter $\mathcal{R}$, the angular scale of the sound horizon at last scattering $l_A$, and the baryon density $\omega_b$. In Fig. \ref{cmbdta} we compare the theoretical results of $l_A$ from the three models at hand against the observations from the Planck18 dataset. The Fig. shows that both DUDF and UDF models fit the observational value.

\subsection{Effective Equation of State parameter}
In standard cosmology, the universe is isotropic and filled with a perfect fluid. The equation of state (EoS) describing this fluid is characterized by a constant parameter $\omega$ that is the ratio of fluid pressure to fluid density. For the $\Lambda$CDM model, today's universe is characterized mainly by two components: dark energy with $\omega = -1$ and dark matter with $\omega = 0$.

\begin{figure} [h!]
	\centering  
		\includegraphics[width=0.6 \textwidth] {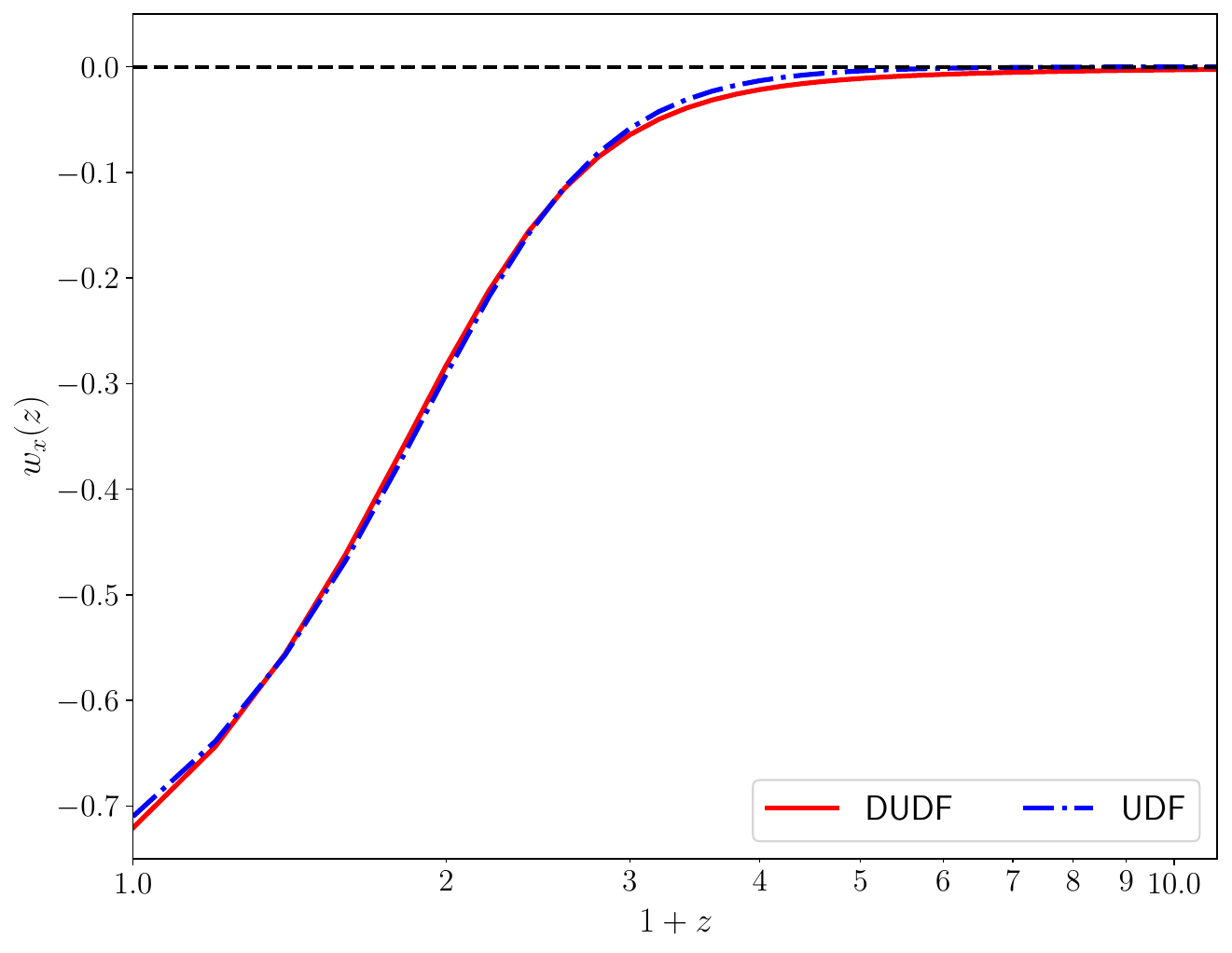}
	\caption{Behaviour of the EoS of fluid models.}
	\label {eosfld}
\end{figure}

On the other hand, the effective EoS parameter of DUDF model is calculated using relations (\ref{pars}) and (\ref{delta}) in (\ref{eqos}) where we get
\begin{equation}
    p_{\rm fld} = - \rho_{\rm fld} + \frac{\rho_{\rm fld}^{m+1}}{{\tilde{\delta}}^m + \rho^{m}_{\rm fld}} - 3 \zeta_0 H .
\label{eqos1}
\end{equation}
Accordingly
\begin{equation}
    \omega_{\rm fld} = - 1 + \frac{\rho_{\rm fld}^{m}}{{\tilde{\delta}}^m + \rho_{\rm fld}^m} - 3 \zeta_0 \frac{H}{\rho_{\rm fld}} .
\label{eqos2}
\end{equation}
In Fig. \ref{eosfld} we present the behavior of the EoS of DUDF and UDF models.
The Fig shows the ability of the fluid for a smooth transition from dust at early times, where $\omega(z) = 0$, to dark energy domination at late time.

\subsection{Deceleration Parameter} 
Deceleration parameter is an important parameter in cosmology. Evolution of the expansion of the universe can be discussed through this parameter. Deceleration parameter is defined as
\begin{equation}
    q(z) = - a(z) \frac{\ddot{a}}{\dot{a}^2} .
\label{qodz}
\end{equation}

\begin{figure} [h]
	\centering 
		\includegraphics[width=0.6 \textwidth] {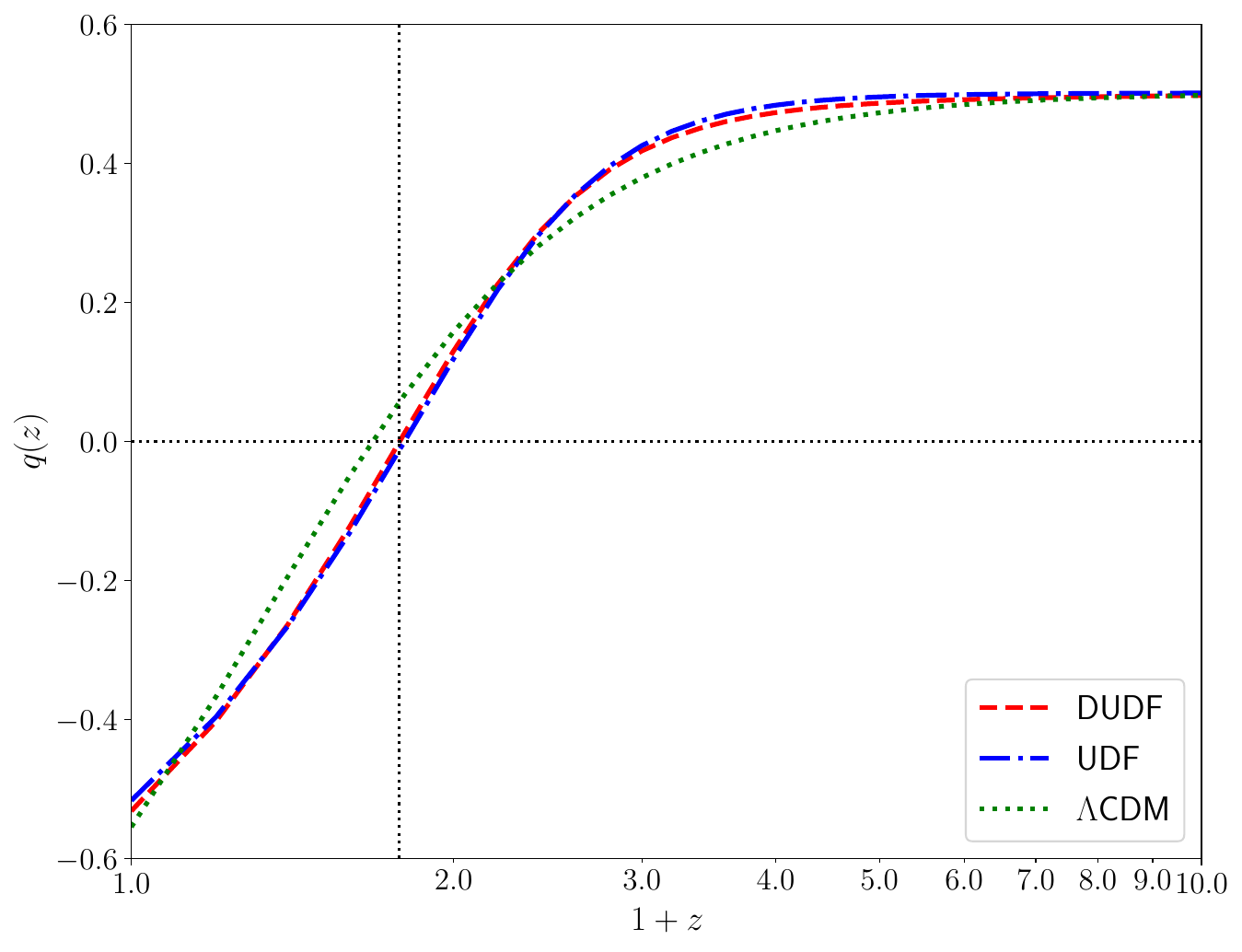}
  	\vspace*{-.4cm}
	\caption{Evolution of deceleration parameter.}
	\label {decn}
\end{figure}

It was believed that the expansion of the universe is decelerating due to self-gravitation. A study of the apparent magnitude of distant supernovae reveals that the universe is instead accelerating. However, cosmological observations indicated that this acceleration is a recent phenomenon, which was explained by the domination of dark energy in the late times. Fig. \ref{decn} displays the behavior of the deceleration parameter due to DUDF and UDF models. The Fig. also shows results from the $\Lambda$CDM scenario for comparison.

One can see from the Fig. that the deceleration parameter due to the fluid model follows the standard evolution of the universe. The fluid can play the role of the cosmological constant in the late times, so able to switch the deceleration expansion to an accelerating one. The Fig also shows that viscosity slows the transition to the accelerating phase. The transition redshift increases from $z_{tr} \approx 0.78$ for DUDF model to $z_{tr} \approx 0.8$ for the UDF where the fluid is non-viscous. These values of $z_{tr}$ are consistent with the results of the observational analysis made in \cite{jes} and \cite{hai}. Similarly, the viscosity affects the current value of the deceleration parameter. Its value increases from $q_{0} \approx -0.53$ for DUDF model to $q_{0} \approx -0.52$ for the UDF model. These values of $q_0$ agree with the results of \cite{varg}.

\subsection{Density of the Universe}
\begin{figure} [h]
	\centering
	\includegraphics[width=0.7 \textwidth] {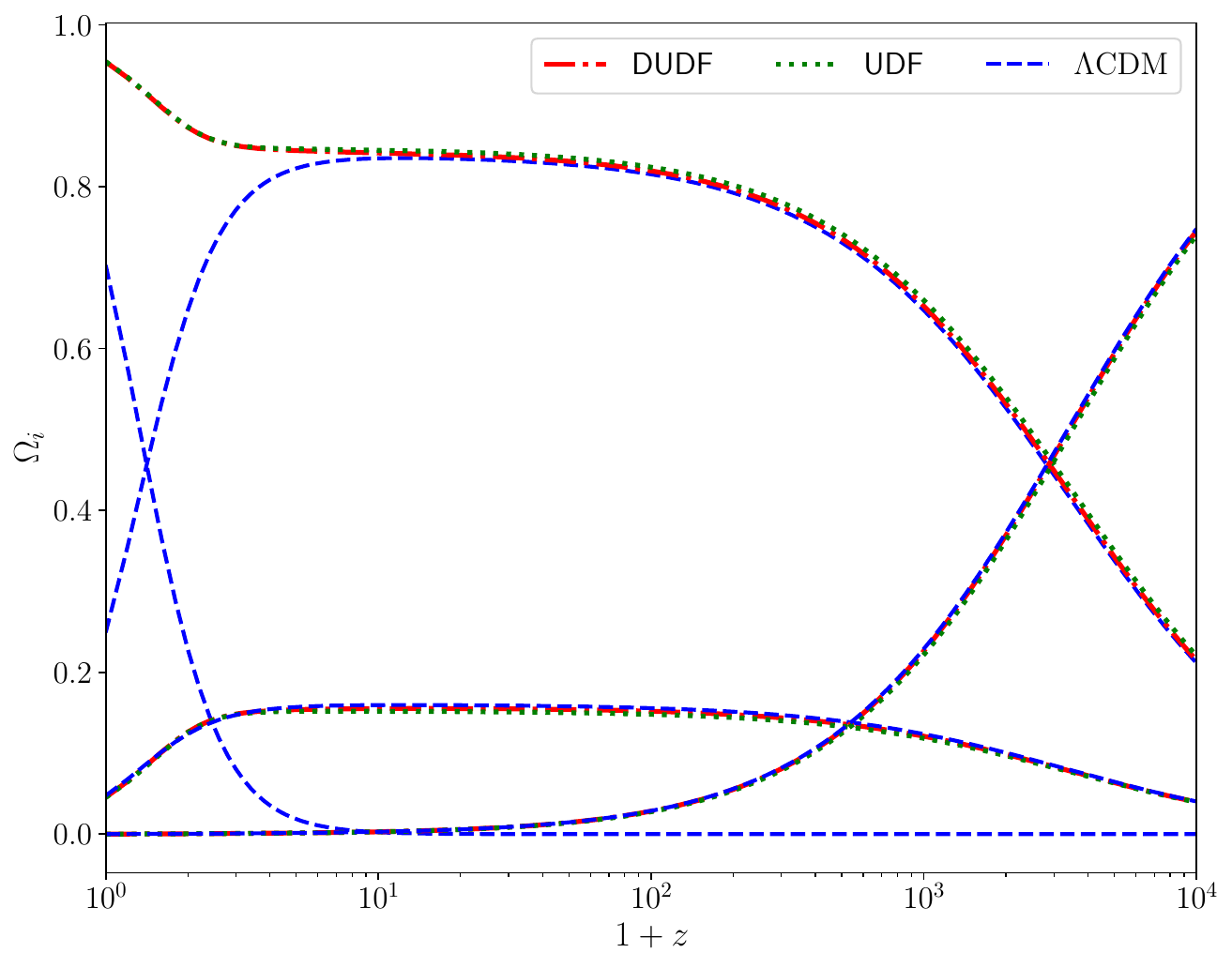}
  	\vspace*{-.4cm}
	\caption{Evolution of density parameter.}
	\label {Omega}
\end{figure}

Observations indicate that the universe is flat with a mean energy density equals to its critical density. Friedmann equations enable the determination of the energy budget of the universe at any time. For any component $x$ of the universe's fluid, the energy density is given by 
\begin{equation}
\Omega_{x}(z)= \frac{\rho_x(z)}{\rho_{crit}(z)} ,
\label{denx}
\end{equation}
where $\rho_x$ is the density of the $x$-component, and $\rho_{crit}$ is the critical density given by
\begin{equation}
\rho_{crit}(z)= \frac{3 c H^2(z)}{8 \pi G} .
\label{denc}
\end{equation}
In Fig. \ref{Omega} we displayed the density of different components of the fluid model vs the $\Lambda$CDM model.

We can see from the Fig. that the density parameter of the unified fluid matches the dust density due to the $\Lambda$CDM scenario at early times while gradually following the dark energy density to the future. 

\section{Estimation of the Viscosity of the Cosmic Fluid}
Different authors estimated the viscosity of the cosmic fluid to be $\zeta_0 \leq 10^7$ $Pa$.$s$. However, its value is highly model-dependent. As we mentioned above, the best model due to the Akaike criteria is the perfect fluid model, UDF, though the result of $\Delta AIC$ for the viscous model DUDF indicated that this model is substantial. We can consider the mean value of $\zeta_0$ due to this model as an estimated value for the viscosity of the cosmic fluid. From table \ref{tabmeans} we see that the mean value of $\zeta_0$ is $2.9$ $km$ $s^{-1}  Mpc^{-1}$. Multiplying this by $c^2/ 8 \pi G $, we get an estimated value for the viscosity of the cosmic fluid as $5.04 \times 10^6$ $Pa$.$s$.

\section{Conclusions}
We considered a standard FRW flat cosmology with several energy source components. The two dark sectors of the cosmic fluid are replaced by a dissipative unified dark fluid. The EoS of this fluid can asymptote between two power laws. As a result, it enables fluid to smoothly evolve between early and late times.  The dissipative effects are described by a bulk viscosity with a constant coefficient.

We have implemented data analysis for the DUDF model as well as its special perfect fluid case, UDF, using recent observational datasets from R20, SNe Ia, OHD, BAO, and CMB. Similar analysis is also performed on the $\Lambda$CDM model for comparison. We found that between the three analyzed models, the DUDF model has
the lower $\chi_{\rm min}^{2}$-value. We then relied on model selection statistics in the form of AIC to compare different models. We found that the UDF model has the minimum AIC value, with the conclusion that the data favored the perfect unified fluid model. On measuring $\Delta_i$ to the lowest $AIC$ model, the UDF model, we found that the DUDF model has a difference of less than 2, meaning that the DUDF model is substantial on the level of empirical support and is not worth more than a bare mention of evidence against the UDF model. Given this result, we estimated a present-day value for the viscosity of the cosmic fluid as $5.04 \times 10^6$ $\rm Pa.s$.

Later on, we studied the evolution of the cosmic fluid through the Hubble parameter, the distance modulus, BAO data, acoustic scale at last scattering $l_A$, the effective equation of state of the dark fluid, the cosmic deceleration parameter, and the cosmic density parameter. Such a study demonstrated that the fluid model highly matches observations and follows the standard scenario throughout the evolution history. The evolution of the deceleration parameter proved to reflect the well-known behavior of such parameter, with a present-day value that agrees with the results obtained by many previous authors. The evolution of the EoS of the fluid manifests its ability for a smooth transition between dark sectors at early and late times. The density parameter indicates again that the fluid can smoothly evolve from one dark sector to the other, with a dust density that matches the $\Lambda$CDM model during the high redshifts, and a dark energy density dominates the future. 

Finally, we conclude that our unified fluid model provides a viable framework for describing the universe.

\section*{Acknowledgments}
The authors would like to thank A. Awad, A. El-zant, and W. El Hanafy for valuable discussions.


\begin{thebibliography}{99}

\bibitem {riss} A.G. Riess, A.V. Filippenko, P. Challis, et al.,  {\it AJ} {\bf116} (1998) 1009–1038.

\bibitem {per} S. Perlmutter, G. Aldering, G. Goldhaber, et al., {\it APJ} {\bf517} (1999) 565–586.

\bibitem {ser1} D.N. Spergel, R. Bean and O. Doré, et al., {\it ApJS} {\bf170} (2007) 377.

\bibitem {dos1} A. Blanchard, M. Douspis, M. Rowan-Robinson and S. Sarkar, {\it A\&A} {\bf449} (2006) 925.

\bibitem {wien} S. Wienberg, {\it Rev. of mod. phys.} {\bf61} (1989) 1.

\bibitem {plnk18} Planck Collaboration, Planck 2018 results VI, N. Aghanim et al., {\it A\&A} {\bf641} (2020) A6.

\bibitem {book} M.F. Wondrak, The Cosmological Constant and Its Problems: A Review of Gravitational Aether, {\it Experimental Search for Quantum Gravity} Hossenfelder S. (eds) FIAS Interdisciplinary Science Series. Springer, Cham (2018).

\bibitem {ombr} L. ombriser, {\it Physics Letters B} {\bf 797} (2019) 134804.

\bibitem{coprob} S.M. Carroll, {\it Living Rev. Rel.} {\bf 4} (2001) 1.

\bibitem{khur} M. Khurshudyan and A.Z. Khurshudyan, {\it Symmetry} {\bf 10} (2018) 577.

\bibitem{pali} A. Paliathanasis, S. Pan and W. Yang, {\it Int. J. mod. phys. D} {\bf 28} 12 (2019) 1950161.

\bibitem {gzaho} Gong-Bo Zhao, et al., {\it Nature Astron.}  {\bf1}, 9 (2017) 627.

\bibitem {revera} C. Escamilla-Rivera and A. Nájera,  arXiv:2103.02097 [gr-qc].

\bibitem{bosh} K. Boshkayev, R. D’Agostino and O. Luongo, {\it Eur. Phys. J. C} {\bf 79} (2019) 332.

\bibitem{wzim} W. Zimdahl, H.E.S. Velten and W.S. Hipolito {\it Int. J. mod. phys. Conf. Ser.} {\bf 03} (2011) 312.

\bibitem {chap1} A. Kamenshchikab, U. Moschellac and V. Pasquier, {\it Physics Letters B} {\bf511} (2001) 265.

\bibitem {chap2} B.R. Dinda, S. Kumar and A.A. Sen, {\it Phys. Rev. D} {\bf90} (2014) 083515.

\bibitem {viab} E. Elkhateeb, {\it Astrophys. Space Sci.} {\bf363} (2018) 7.

\bibitem {zel} Ya B Zel'dovich {\it Sov. Phys. Usp.} {\bf9} (1967) 602.

\bibitem{mis}  C. W. Misner {\it Phys. Rev. Lett.} {\bf19} (1967) 533.

\bibitem{wein} S. Weinberg {\it Apj} {\bf168} (1971) 175.

\bibitem{bel75} V.A. Belinskil and I.M. Khalatnikov {\it Zh. Eksp. Teor. Fiz.} {\bf 69} (1975) 401.

\bibitem{bel78}  V.A. Belinskil and I.M. Khalatnikov {\it Zh. Eksp. Teor. Fiz.} {\bf 72} (1977) 3.

\bibitem{lima} J.A.S. Lima, A.S.M. Germano {\it Phys. Lett. A} {\bf 170} (1992) 373.

\bibitem{pav} W. Zimdahl and  D. Pavon {\it  Gen. Relativ. Gravit.} {\it 26} (1994) 1259.

\bibitem{zim96} W. Zimdahl {\it Phys. Rev. D} {\bf 53} (1996) 5483.

\bibitem{zim97} W. Zimdahl {\it Mon. Not. R. Astron. Soc.} {\bf 288} 665.

\bibitem{fran} J.S. Farnes {\it A \& A} {\bf 620} (2018) A92.

\bibitem{giov} M. Giovannini {\it Physical Review D} {\bf 93} (2016) 083521.

\bibitem {ent1} W.S. Hipólito-Ricaldi, H.E.S. Velten and W. Zimdahl, {\it JCAP} {\bf0906} (2009) 016.

\bibitem {ent2} J.C. Fabris, P.L.C. de Oliveira and H.E.S. Velten, {\it Eur. Phys. J. C} {\bf71} (2011) 1773.

\bibitem {ani} I. Brevik, {\o}. Gr{\o}n {\it Journal of Magnetohydrodynamics and Plasma Research} {\bf 19}, Iss. 1/2 (2014) 97.

\bibitem{inf1} S. Nojiri, S.D. Odintsov, V.K. Oikonomou and T. Paul {\it Phys. rev. D} {\it 102} (2020) 023540.

\bibitem{brti} I. Brevik and A.V. Timoshkin {\it JETP} {\bf 122} (2016) 679.
 
\bibitem{kbamba} K. Bamba and S.D. Odintsov {\it Eur. Phys. J. C} {\it 76} (2016) 18.

\bibitem{camp} S. Del Campo, R. Herrera and D. Pavon {\it Phys. Rev. D} {\it 75} (2007) 083518.

\bibitem{grn} {\o}. Gr{\o}n {\it Astrophys. Space Sci.} {\bf 173} (1990) 191.

\bibitem{iov} I. Waga, R.C. Falcão, and R. Chanda {\it Phys. Rev. D} {\bf 33} (1986) 1839.

\bibitem{inf2} L. Diosi, B. Keszthelyi, B. Lukacs and G. Paal {\it Phys. Lett.} {\bf 157B} (1985), 1, 23.

\bibitem{late1} I. Brevik and A.V. Timoshkin (2020) arXiv:2011.09207 

\bibitem{mosas} N.D.J. Mohana, A. Sasidharanb and T. K Mathewc {\it Eur. Phys. J. C} {\bf 77} (2017) 849.

\bibitem{ibrevik} I. Brevik, {\o}. Gr{\o}n, J. de Haro, S.D. Odintsov and E.N. Saridakis {\it Int. J. Mod. Phys. D} {\bf 26} (2017) 1730024.

\bibitem{ibrev} I. Brevik {\it Mod. Phys. Lett. A} {\bf 31} (2016) 8.

\bibitem{late2} I. Brevik, E. Elizalde, S. Nojiri, and S.D. Odintsov {\it Phys. Rev. D} {\bf 84} (2011) 103508.

\bibitem{sym} I. Brevik and B.D. Normann {\it Symmetry} {\bf 12} (2020) 1085.

\bibitem{vic} V.H. C$\acute{\texttt{a}}$rdenas, M. Cruz and S. Lepe {\it Phys. Rev. D} {\bf 102} (2020) 123543.

\bibitem{ten1} E. Elizalde, M. Khurshudyan, S.D. Odintsov, and R. Myrzakulov {\it Phys. Rev. D} {\bf 102} (2020) 123501.

\bibitem{silva} W.J.C da Silva and R. Silva {\it Eur. Phys. J. C} {\bf 81} (2021)  403.

\bibitem{ten2} D. Wang, Y.J Yan, and X.H Meng {\it Eur. Phys. J. C} {\bf 77} (2017) 660.

\bibitem{uni1} I. Brevik and A.V. Timoshkin {\it International Journal of Geometric Methods in Modern Physics} {\bf 17}, 2 (2020) 2050023.

\bibitem{wyang} W. Yang, S. Pan, y. E. Di Valentino, z. A. Paliathanasis and J. Lu {\it Phys. Rev. D} {\bf 100} (2019) 103518.

\bibitem {diss4} E. Elkhateeb {\it Int. J. mod. phys. D} {\bf28}, no. 9 (2019) 1950110.

\bibitem{strip} S.K. Tripathy, D. Behera and B. Mishra {\it Eur. Phys. J. C} {\bf 75} (2015) 149.

\bibitem{folomeev} V. Folomeev and V. Gurovich {\it Phys. Lett. B} {\bf 661} Issues 2–3  (2008) 75.

\bibitem{uni2} J. Ren and X.H. Meng {\it Int. J. Mod. Phys. D} {\bf 16}, 08 (2007) 1341.

\bibitem {diss1} N. Cruz1, E. Gonz$\acute{\texttt{a}}$lez, S. Lepe and D. S$\acute{\texttt{a}}$ez-Chill$\acute{\texttt{o}}$n G$\acute{\texttt{o}}$mez {\it JCAP} {\bf 12} (2018) 017.

\bibitem {sanand} S. Anand, P. Chaubal, A.M. azumdar and S.M. ohanty {\it JCAP} {\bf 11} (2017) 005.

\bibitem {Miguel} M. Cruz, N. Cruz and S. Lepe {\it Phys. Rev. D} {\bf 96} (2017) 124020.

\bibitem {ncomat} N. Komatsu and S. Kimura {\it Phys. Rev. D} {\bf 90} (2014)
 123516.

\bibitem {diss2} A. Avelino and U. Nucamendi, {\it JCAP} {\bf2010}, no. 08 (2010) 009.

\bibitem {diss3} J.C. Fabris, S.V.B. Goncalves and R. de Sa Ribeiro, {\it Gen.Rel.Grav.} {\bf38} (2006) 495.

\bibitem {bing} Bing Xu et al., {\it ApJ} {\bf855} (2018) 89.

\bibitem {neff} P. F. de Salas and S. Pastor, {\it J. Cosmology Astropart. Phys.} {\bf07} (2016) 051.

\bibitem {pan} D.M. Scolnic et al., {\it ApJ} {\bf859} (2018) 101.

\bibitem {cctec}  M. Moresco et al., {\it JCAP} {\bf05} (2016) 014.   

\bibitem {chron} R. Ji$\acute{\texttt{m}}$enez and A. Loeb, {\it Astrophys. J.} {\bf573} (2002) 37. 

\bibitem {mag} J. Maga$\widetilde{n}$a, M.H. Amante, M.A. Garcia-Aspeitia, V. Motta, {\it Mon. Not. Roy. Astron. Soc.} {\bf476} (2018) 1, 1036-1049.

\bibitem{alam} S. Alam et al., {\it Mon. Not. R. Astron. Soc.} {\bf470} (2017) 2617.

\bibitem{ries20} A.G. Riess, S. Casertano, W. Yuan, J.B. Bowers, L. Macri, J.C. Zinn and D. Scolnic {\it Astrophys. J. Lett.} {\bf908} (2021) L6.

\bibitem{aiz} Andoni Aizpuru, Rubén Arjona, and Savvas Nesseris {\it Phys. Rev. D} {\bf104} (2021) 043521.

\bibitem{data} https://sdss3.org/science/boss\_publications.php.

\bibitem{chen}  Lu Chen et al., {\it JCAP} {\bf02} (2019) 028.

\bibitem{mcmc} D.W. Hogg and D. Foreman-Mackey {\it The Astrophysical Journal Supplement Series} {\bf236} (2018) 11.

\bibitem{Akke} H. Akaike {\it IEEE Transactions on Automatic Control} {\bf19} (1974) 716.

\bibitem{park} C.G. Park {\it Physical Review D} {\bf91}, Issue 12 (2015) 123519.

\bibitem{Burn} K.P. Burnham, D.R. Anderson {\it Model Selection and Multimodel Inference: A Practical Information-Theoretical Approach} (2002) Second edition, Springer, New York.

\bibitem{jes} J.F. Jesus, R.F.L. Holanda and S.H. Pereira {\it JCAP} {\bf05} (2018) 073.

\bibitem{hai} Hai Yu, et al {\it Apj} {\bf856} (2018) 3.

\bibitem{varg} M. Vargas dos Santos et al {\it JCAB} {\bf2016}, Issue 02 (2016) 066.

\end{thebibliography}
\end{document}